\documentclass[a4paper,preprint]{emulateapj}
\usepackage{amstext}
\usepackage{apjfonts}
\usepackage{amsmath}

\begin{document}
\title{Radio Imaging of GRB Jets in Nearby Supernovae}

\author{Jonathan Granot and Abraham Loeb \altaffilmark{1}}

\affil{Institute for Advanced Study, Princeton, NJ 08540; 
granot@ias.edu, loeb@ias.edu}

\altaffiltext{1}{Guggenheim Fellow; on sabbatical leave from the
Astronomy Department, Harvard University, Cambridge, MA 02138}

\begin{abstract}

We calculate the time evolution of the flux, apparent size, and image
centroid motion of gamma-ray burst (GRB) radio jets, and show
that they can be resolved by the VLBA at distances of hundreds of
Mpc. We find that GRB 030329 which showed spectroscopic evidence for
an associated Type Ic supernova (SN) at a distance of $\approx
800\;$Mpc might just be resolvable by VLBA after several months.  The
prospects are much better for jets that are oriented sideways in
similar SNe with no GRB counterpart; in particular, the motion of the
flux centroid in such jets can be detected by VLBA up to $z\sim 1$
even when the jet cannot be resolved.  If most GRBs are accompanied by
a SN Ib/c, then there should be a few SN/GRB jets per year within a
distance $\lesssim 200\;$Mpc and most of them would be oriented
sideways with no $\gamma$-ray or X-ray precursor.  Detection of these
jets can be used to calibrate the fraction of all core collapse SNe
that produce relativistic outflows and determine the local GRB rate.
Overall, the rate of SNe Ib/c which do not produce a GRB at all, but
rather make relativistic radio jets with an initial Lorentz factor of
a few, may be larger by up to two orders of magnitude than the rate of
those that produce GRBs.

\end{abstract}


\section{Introduction}

Recent evidence indicates that long-duration $\gamma$-ray bursts (GRBs)
are associated with Type Ic supernovae (SNe); of particular
significance is the 3500\,--\,8500\,\AA\ spectrum of SN 2003dh
associated with GRB 030329 (Stanek et al. 2003), which was very
similar to that of SN 1998bw/GRB 020405 (Nakamura et al. 2001).  This
supports previous, more circumstantial evidence, such as late time
bumps in afterglow lightcurves (Bloom 2003) and the association of
GRBs with central star-forming regions of galaxies (Bloom, Kulkarni,
\& Djorgovski 2002).  This evidence raises two basic questions: (1)
which fraction of all core collapse SNe produce relativistic
outflows?, and (2) what is the probability distribution of the
collimation angle, initial Lorentz factor, and energy output of these
outflows?

For every nearby ($\lesssim 1\;$Gpc) event like GRB 030329, there
should be hundreds of GRB jets that are not pointed at us, based on
existing estimates for the jet opening angle (Frail et al. 2001). As
sufficiently close GRBs can be resolved by radio telescopes (Woods \&
Loeb 1999; Cen 1999; Granot, Piran \& Sari 1999; Ayal \& Piran 2001;
Paczynski 2001), an effective method to address the above two
questions is to search for relativistic outflows in nearby core
collapse SNe. In \S \ref{calc} we calculate the expected flux,
apparent size, and image centroid motion of semi-relativistic GRB
radio jets viewed sideways. Such jets would have no observable GRB
precursor but can be identified and timed via their associated SN
emission.  The early images of relativistic SN/GRB radio jets may
resemble relativistic radio jets in quasars (Begelman, Blandford \&
Rees 1984) or micro-quasars (Mirabel \& Rodriguez 1999).  However,
while quasars often inject energy continuously into the jet, GRB
sources are impulsive.  Although quasar jets remain highly collimated
throughout their lifetimes, GRB jets decelerate and expand
significantly once they become non-relativistic, $\sim 1\;$yr after
the explosion. The hydrodynamic remnant of a GRB eventually becomes
nearly spherical after $\sim 5\times 10^3\;$yr (Ayal \& Piran 2001).

\section{Flux, Size, and Centroid Shift, of GRB Jets}
\label{calc}

First, we calculate the radio flux from GRB jets observed from
different viewing angles, $\theta_{\rm obs}$, with respect to to the
jet axis. We assume a double-sided jet, and calculate the emission
from both the forward jet that points towards the observer
($\theta_{\rm obs,1}\le\pi/2$), and the opposite counter-jet
($\theta_{\rm obs,2}=\pi-\theta_{\rm obs,1}\ge\pi/2$).  Off-axis
lightcurves from GRB jets were already calculated using different
models with various degrees of complexity (Granot et al. 2002),
varying from simple models (Dalal, Griest \& Pruet 2002; Rossi,
Lazzati \& Rees 2002) to numerical simulations (Granot et al. 2001).
Compared to simpler models, simulations typically show differences of
order unity in the flux around or after the time of the peak in the
light curve, a much more moderate rise in the flux before the peak,
and a much smoother peak at $\theta_{\rm obs}\lesssim 3\theta_0$,
where $\theta_0$ is the initial jet opening angle.  The more moderate
rise before the peak is due to relatively slow material at the sides
of the jet which emits much more isotropically compared to the front
of the jet ($\theta\lesssim\theta_0$) where the emission is strongly
beamed away from off-axis observers at early times.  A simple model
for the emission from the material behind the bow shock of GRB jets,
which essentially addresses the same emission component, was also
investigated by Wang \& Loeb (2001).

For simplicity, we adopt an extended version of model 1 from Granot et
al. (2002); where appropriate, we mention the qualitative differences
that are expected in more realistic jet models.  We assume a point
source that moves along the jet axis. Its radial location $R$, the lab
frame time $t$, and the observed time $t_{\rm obs}$ are related by
$t_{\rm obs}=t-R\cos\theta_{\rm obs}$.  For an on-axis observer
($\theta_{\rm obs}=0$) we assume a broken power law spectrum (Sari,
Piran \& Narayan 1998). The values of the peak flux and break
frequencies before the jet break time $t_j$ ($t_{{\rm obs},j}$) are
taken from Granot \& Sari (2002).  t $t_j<t<t_{\rm NR}$, where $t_{\rm
  NR}$ ($t_{\rm obs,NR}$) is the non-relativistic transition time, the
temporal scaling of the peak flux and break frequencies is modified
according to Rhoads (1999) and Sari, Piran \& Halpern (1999).  At
$t>t_{\rm NR}$ the scalings are changed to those for the Sedov-Taylor
regime (e.g. Frail, Waxman, \& Kulkarni 2000).  The light curve for
off-axis observers is then calculated using the appropriate
transformation of the radiation field, $F_\nu[\theta_{\rm obs},t_{\rm
  obs}(\theta_{\rm obs})]= a^3F_{\nu/a}[0,t_{\rm obs}(0)]$, where
$a=(1-\beta)/(1-\beta\cos\theta_{\rm obs})$.  In Granot et al. (2002),
it was also assumed that $t_{\rm obs}(0)/t_{\rm obs}(\theta_{\rm
  obs})=a$, which is an approximation since actually $dt_{\rm
  obs}(0)/dt_{\rm obs}(\theta_{\rm obs})=a$, and is not very accurate
for $t_j<t<t_{\rm NR}$ when $\gamma$ drops exponentially with radius.
In this case, we must use the more accurate relation, $t_{\rm
  obs}=t-R\cos\theta_{\rm obs}$, where
$R=\int_0^t\beta(\tilde{t})d\tilde{t}$.  The Lorentz factor is
approximately given by $\gamma\approx\theta_0^{-1}(t/t_j)^{-3/2}$ at
$t\leq t_j$, $\gamma\approx\theta_0^{-1}\exp(1-t/t_j)$ at $t_j\le t\le
t_{\rm NR}$ and $\gamma\approx\left[1-\beta_{\rm NR}^2 (t/t_{\rm
    NR})^{-6/5}\right]^{-1/2}$ at $t\ge t_{\rm NR}$, where we use
$\beta_{\rm NR}=\beta(t_{\rm NR})\equiv 0.5$ ($\gamma_{\rm
  NR}=2/\sqrt{3}\approx 1.15$) for the transition to the
non-relativistic regime.  We find
\begin{equation}\label{R}
R\approx\left\{\begin{matrix}
[1-(t/t_j)^{3}\theta_0^2/16]ct & \ \ t\le t_j\vspace{0.1cm}\cr
\{1-[\exp(2t/t_j-2)-1/2](\theta_0^2t_j/8t)\}ct & \ \ t_j\le t\le t_{\rm NR}
\vspace{0.1cm}\cr 
R_{\rm NR}+[(t/t_{\rm NR})^{2/5}-1](5/2)\beta_{\rm NR}ct_{\rm NR}
& \ \ t\ge t_{\rm NR}\end{matrix}\right.\ ,
\end{equation}
where $R_{\rm NR}=R(t_{\rm NR})$. Following Sari et al.  (1999), we
use $t_{\rm obs}(\theta_{\rm obs}=0)=R/4\gamma^2c$ instead of
$R/16\gamma^2c$, and adopt their expression for $t_{\rm obs,j}(0)$ and
the relation $t_j=4t_{\rm obs,j}/\theta_0^2$.  Using a simple energy
equation, $E_{\rm iso}=2E/\theta_0^2=(4\pi/3)nm_pc^2R^3\gamma^2$, and
$t_{\rm NR}/t_j=1-\ln x$ where $x\equiv \gamma_{\rm NR}\theta_0$, we
obtain\footnote{$f\approx 3$ for $\theta_0=0.1$ and has only a weak
  logarithmic dependence on $\theta_0$.}
\begin{equation}\label{R_j}
R_j\equiv R(t_j)=\left(\frac{3E}{2\pi n m_pc^2}\right)^{1/3}
=6.8\times 10^{17}\left(\frac{E_{51}}{n_0}\right)^{1/3}\;{\rm cm}\ ,
\end{equation}
\begin{equation}\label{R_NR}
\frac{R_{\rm NR}}{R_j}\equiv f=
\frac{1-\ln x-[x^{-2}-1/2]\theta_0^2/8}{1-\theta_0^2/16}
\approx 1-\ln x-\frac{1}{8\gamma_{\rm NR}^2}\ ,
\end{equation}
where $n=n_0\;{\rm cm^{-3}}$ is the external density and
$E=10^{51}E_{51}\;$erg is the energy in the two jets.

\begin{figure}
\plotone{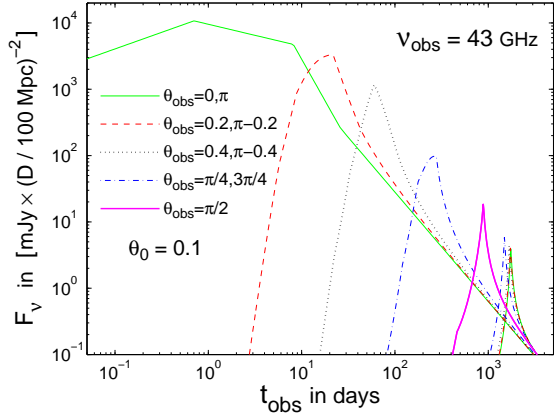}
\caption{\label{fig1} Radio light curves
($\nu_{\rm obs}=43\;$GHz) for different viewing angles, and
$E=10^{51}\;$erg, $n=1\;{\rm cm^{-3}}$, $\theta_0=0.1$,
$\epsilon_e=0.1$, $\epsilon_B=0.01$, $p=2.5$.  The jet that points
towards us ($\theta_{\rm obs,1}\leq\pi/2$) and the counter jet that
point away from us ($\theta_{\rm obs,2}=\pi-\theta_{\rm
obs,1}\geq\pi/2$) are shown using the same line style (the
emission from the counter jet peaks at a later time).  The SN radio
emission is expected to be much weaker in the plotted range of
observed times.}
\end{figure}

Figure \ref{fig1} shows the radio light curves at different
$\theta_{\rm obs}$, for both the forward and counter jets. 
While the optical is typically above the peak frequency
$\nu_m$ at $t_j$, the radio is typically below $\nu_m$, and
therefore the temporal decay slope $\alpha=-d\ln F_\nu/d\ln t_{\rm
obs}$ for $\theta_{\rm obs}<\theta_0$ changes from $-1/2$ to $1/3$ at $t_{{\rm
obs},j}$. After $t_{{\rm obs},m}$ when $\nu_m$ sweeps past the
observed frequency $\nu_{\rm obs}$, $\alpha=p$, while at $t>t_{\rm
NR}$ $\alpha=(15p-21)/10$, where $p\sim2-2.5$ is the
power-law index of the electron energy distribution.  
For $\theta_{\rm obs}>\theta_0$ the flux still
rises at $t_{{\rm obs},j}<t_{\rm obs}<t_{{\rm obs},m}$; for
$\theta_{\rm obs}\lesssim\pi/4$ the flux peaks at $t_{{\rm
obs},p}=t_{{\rm obs},m}$, while for $\theta_{\rm obs}\gtrsim\pi/4$,
$t_p=t_{\rm NR}$. If $t_{{\rm obs},p}=t_{{\rm obs},m}$, the spectral 
slope should change from $F_\nu\propto\nu^{1/3}$ to $\nu^{(1-p)/2}$ at
$t_{{\rm obs},p}$, which should be easy to observe. If $t_p=t_{\rm NR}$, 
then $t_{{\rm obs},p}\sim t_{{\rm obs},j}(\theta_{\rm obs}/\theta_0)^2
\sim t_{\rm NR}\theta_{\rm obs}^2/4$ for $\theta_0<\theta_{\rm obs}\ll 1$,
while for the counter jet $t_{{\rm obs},p}\sim t_{\rm NR}$,
so that the ratio of the two peak times is 
$\sim 4/\theta_{\rm obs}^2$, and may be used to estimate 
$\theta_{\rm obs}$. For $\theta_{\rm obs}\gtrsim\pi/4$, $t_{{\rm obs},p}$
for the two jets is less than a factor of $\sim 10$ apart.
The light curves for the counter
jets with $\theta_{\rm obs,2}\gtrsim 3\pi/4$ 
($\theta_{\rm obs,1}\lesssim\pi/4$) are all very similar, and peak at
$\sim 2R_{\rm NR}/c\approx 2t_{\rm NR}$. 
For $\theta_{\rm obs}>\theta_0$ the peak flux is a factor of a few
larger than the flux for $\theta_{\rm obs}=0$ at the same $t_{\rm obs}$.
For more realistic jet models we expect
a much smoother peak and a somewhat smaller peak flux.
We note that $F_\nu(t_{\rm NR})\propto E$ (Frail, Waxman \& Kulkarni 2000), 
so that a larger energy implies a larger flux at $t_{\rm NR}$.
For example, keeping the same energy per solid angle, and increasing
$\theta_0$ from $0.1$ to $0.3$ ($0.5$), would increase $E$ and
therefore $F_\nu(t_{\rm NR})$ by a factor of $9$ ($25$), and would 
modify the afterglow light curves (Granot et al. 2002).

\begin{figure}
\plotone{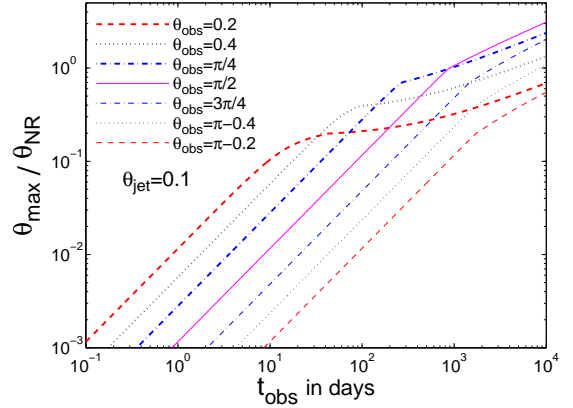}
\caption{\label{fig2} The
angular extent of the jet, $\theta_{\rm max}$, in units of 
$\theta_{\rm NR}$ (e.g. Eq. \ref{theta_NR}), for the same viewing 
angles and jet parameters as in Fig. \ref{fig1}.}
\end{figure}

Figure \ref{fig2} shows the maximal angular size\footnote{If both the
forward and counter jets are visible, the total angular 
size would be the sum their two $\theta_{\rm max}$.}
$\theta_{\rm max}$, 
of the jets and counter jets from Fig. \ref{fig1}, in units of 
\begin{equation}\label{theta_NR}
\theta_{\rm NR}=\frac{R_{\rm NR}}{D}=1.4\left(\frac{f}{3}\right)
\left(\frac{E_{51}}{n_0}\right)^{1/3}
\left(\frac{D}{100\;{\rm Mpc}}\right)^{-1}\;{\rm mas}\ ,
\end{equation}
measured from the center.  The VLBA has an angular resolution
of\footnote{ see
  http://www.aoc.nrao.edu/vlba/obstatus/obssum/node30.html} $\sim
170\;\mu$as at $43\;$GHz, and may resolve the jet around $t_{\rm NR}$
(typically a few months to years after the SN), up to distances of
$D\sim 1\;$Gpc.  The expected peak flux at $43\;$GHz for a jet with
$\theta_{\rm obs}\sim\pi/4$ at $D\sim 1\;$Gpc is $\sim 1\;$mJy.

\begin{figure}
\plotone{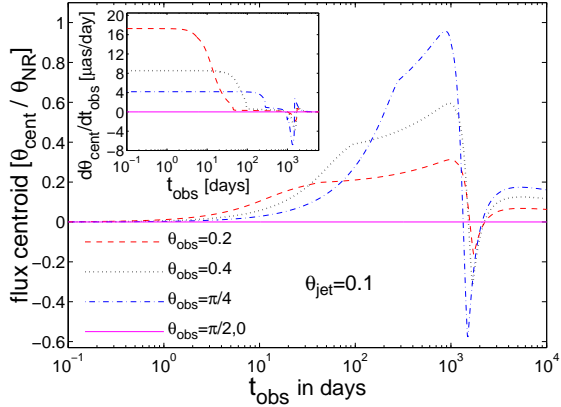}
\caption{\label{fig3} The 
angle of the flux centroid relative to the center of the SN 
explosion, $\theta_{\rm cent}$, along the projection of the
forward jet (with $\theta_{\rm obs}<\pi/2$) on the sky,
in units of $\theta_{\rm NR}$ (e.g. Eq. \ref{theta_NR}).
The inset shows $d\theta_{\rm cent}/dt_{\rm obs}$ for $D=100\;$Mpc.}
\end{figure}

The apparent velocity of the source on the sky is $v_{\rm ap}\equiv
(d\theta_{\rm max}/dt_{\rm obs})D\approx R\sin\theta_{\rm
  obs}/(t-R\cos\theta_{\rm obs})$. For $t\ll t_j$, $v_{\rm
  ap}\approx{\rm const}$ (see inset of Fig. \ref{fig3}) and is
$\approx c$ for $\theta_{\rm obs}=\pi/2$, sub-luminal for $\theta_{\rm
  obs}>\pi/2$ and super-luminal for $\theta_{\rm obs}<\pi/2$. For
$\theta_{\rm obs}\ll 1$ we obtain $v_{\rm ap}\sim 2c\theta_{\rm
  obs}/(\theta_{\rm obs}^2+1/8\gamma^2)$ which for $\theta_{\rm
  obs}>1/\gamma$ is $\sim 2c/\theta_{\rm obs}$.  For $\theta_{\rm
  obs}\lesssim 1/\gamma$ we do not obtain the familiar result $v_{\rm
  ap}\sim\gamma c$ for the afterglow image, since we consider a point
source at a fixed angle $\theta$ from our line of sight, while the
edge of an afterglow image viewed on-axis is at $\theta\sim 1/\gamma$,
where $\gamma$ changes with time (substituting $\theta_{\rm obs}\sim
1/\gamma$ in our formula reproduces this result). However, we are
mainly interested in $\theta_{\rm obs}>\theta_0$, for which our simple
estimate of the source size is reasonable.  When $\gamma$ becomes
$\lesssim 1/\theta_{\rm obs}$, $v_{\rm ap}$ begins to decrease. At
this stage, if the jet expands sideways significantly (i.e.
$\theta_{\rm jet}\sim 1/\gamma$) and if the emission from the whole
jet is taken into account, then $v_{\rm ap}\sim\gamma c$ and
$\theta_{\rm max}(t_{\rm NR})\sim\theta_{\rm NR}$ for all $\theta_{\rm
  obs}$, and not just for $\theta_{\rm obs}\gtrsim\pi/4$.  Just how
$\theta_{\rm max}(t_{\rm NR})$ changes with $\theta_{\rm obs}$,
depends on how close to spherical the jet is at $t_{\rm NR}$.
Numerical simulations show that the jet does not expand laterally very
much before $t_{\rm NR}$ and may approach spherical symmetry only long
after $t_{\rm NR}$ (Granot et al. 2001; Ayal \& Piran 2001).

Figure \ref{fig3} shows the angular location of the flux centroid,
$\theta_{\rm cent}$.  Our results are consistent with those of Sari
(1999) for $\theta_{\rm obs}<\theta_0$ and $t<t_{\rm NR}$.  The inset
shows $d\theta_{\rm cent}/dt_{\rm obs}=v_{\rm ap}/D$ for $D=100\;$Mpc.
Due to symmetry, $\theta_{\rm cent}\equiv 0$ for $\theta_{\rm
  obs}=0,\pi/2$.  At $t<t_{\rm NR}$ the forward jet is much brighter
than the counter jet for $\theta_{\rm obs}\lesssim\pi/4$, and the flux
centroid largely follows the forward jet. However, when the counter
jet peaks at $\sim t_{\rm NR}$, it becomes somewhat brighter than the
forward jet, so that the flux centroid gets closer to the location of
the counter jet (as may be seen from the negative values of
$\theta_{\rm cent}$ in Fig. \ref{fig3}). After $t_{\rm NR}$ the
forward and counter jets have almost the same brightness and the flux
centroid moves very close to the location of the SN, which is midway
between the two jets.  This implies a rather large change in the
location of the flux centroid $\sim\theta_{\rm NR}$ for $\theta_{\rm
  obs}\sim\pi/4$, near $t_{\rm NR}$. For a more realistic jet model,
the peak of the counter jet light curve is expected to be flatter and
at a somewhat lower flux level, so that it is less clear if it will
peak above the emission from the forward jet. Therefore it is not
obvious whether $\theta_{\rm cent}$ will actually obtain negative
values near $t_{\rm NR}$. However, the main conclusion, that a large
change in $\theta_{\rm cent}$ of $\sim\theta_{\rm NR}$ for
$\theta_{\rm obs}\sim\pi/4$, is expected near $t_{\rm NR}$ over a time
scale $\Delta t_{\rm obs}\lesssim t_{\rm obs}$, as well as the
conclusion that $\theta_{\rm cent}$ approaches zero at $t>t_{\rm NR}$,
are robust (unless the two jets are not identical or encounter a
different external density).

The best {\it r.m.s.} error on the localization of the flux centroid
was reported as $10\;\mu$as (Fomalont \& Kopeikin 2003). Such an
accurate localization requires a nearby bright radio quasar on the
sky.  Compared to the best available angular resolution, $\theta_{\rm
cent}$ can be determined with an accuracy better by a factor $\sim
10-20$.  Therefore, the movement of the flux centroid on the sky may
be detected even when the jet is not resolved (i.e. at early times,
$t\ll t_{\rm NR}$ for relatively nearby sources, or near $t_{\rm NR}$
for more distant sources potentially up to cosmological distances,
$z\sim 1$, although such sources would be dim at that age, $\sim
0.1\;$mJy).

At sufficiently late times, $\gtrsim t_{\rm NR}\sim 1\;$yr, when the
jets become non-relativistic and begin to approach a spherical
configuration, one may estimate their physical parameters, i.e.  $E$,
$n$, and the equipartition parameters for the electrons ($\epsilon_e$)
and the magnetic field ($\epsilon_B$), in a similar way as was done
for GRB 970508 by Frail, Waxman \& Kulkarni (2000). This can help
constrain the structure of GRB jets and test if they are uniform or
vary smoothly.

\section{Application for GRB 030329}
\label{030329}

GRB 030329 was detected at a very low redshift of $z=0.1685$ (Greiner
et al. 2003) or an angular sidtance $D\approx 590\;$Mpc.  
Despite its low energy output in
$\gamma$-rays, bumps in its optical afterglow light curve provide
evidence for later energy injection by refreshed shocks which bring
the energy of the afterglow shock close to its average value for all
GRBs (Granot, Nakar \& Piran 2003). Thus we expect $E_{51}\approx 1$.
Since the prompt GRB was observed, $\theta_{\rm obs}\le\theta_0$ and
the emission from the counter jet should peak at $\sim 2t_{\rm
  NR}\approx 2R_{\rm NR}/c\sim 5(E_{51}/n_0)^{1/3}\;$yr, at a flux of
$\sim 20\;\mu$Jy, which would be difficult to detect.  However,
$t_{\rm obs,NR}$ for the forward jet could be somewhat earlier, around
a few months, due to light travel effects and since the jet is still
mildly relativistic at $t_{\rm NR}$.  If the jet spreads sideways
significantly during the relativistic phase ($\theta_{\rm jet}\sim
1/\gamma$), then its angular size after a few months should be
$\sim\theta_{\rm NR}\sim 270(E_{51}/n_0)^{1/3}\;\mu$as, which just might
be resolved by VLBA. However, if the lateral spreading of the jet
during the relativistic stage is modest, the jet might be resolvable
only after a few years when it becomes more spherical but rather dim
(a few tens of $\mu$Jy). The expected shift in the flux centroid from
early times to several months may be up to an angle of
$\sim\sin\theta_{\rm obs}\theta_{\rm NR}\lesssim \theta_0\theta_{\rm
  NR}\approx 19(E_{51}/n_0)^{1/3}\;\mu$as, which might just be
detectable with the VLBA if $\theta_{\rm obs}\approx \theta_0$.

\section{Comparing the rate of Supernovae Ib/c and GRBs}

The rate of Type Ib/c SNe in spiral galaxies is estimated to be $\sim
0.2$ per century per $10^{10} {\rm L_B(\odot)}$ (Prantzos \& Boissier
2003). The luminosity density of the local universe (Glazebrook et al.
2002; Blanton et al. 2002), $\sim 10^8~{\rm L_B(\odot)~Mpc^{-3}}$,
implies a rate density of SN Ib/c of $\sim 2\times 10^4~{\rm
  Gpc^{-3}~yr^{-1}}$.  The collimation-corrected rate of GRBs is
estimated to be (Frail et al. 2001) $\sim 250~{\rm Gpc^{-3}~yr^{-1}}$.
Hence, only $\sim 1\%$ of all SN Ib/c may be associated with
GRBs.\footnote{Norris (2002) has a more optimistic prediction that
  $\gtrsim 25\%$ of all SNe Ib/c produce a sub-class of low-luminosity
  GRBs similar to GRB 980425/SN 1998bw.}

However, more SNe may have relativistic outflows with low Lorentz
factors that would not result in GRBs (which require an initial
Lorentz factor $\Gamma_0\gtrsim 100$), but rather in UV (for
$\Gamma_0\sim 10$) or radio (for $\Gamma_0\lesssim 3$) transients
only. The observational constraints on the rates of such transients
are weak.  Calibration of the statistics of relativistic radio jets in
core collapse SNe can be used to infer the rate of such transients
(which should occur on the rare occasions when the same jets are
viewed on-axis). It can also provide new and more reliable evidence
for the collimation of GRB jets, and an independent estimate for the
distribution of the collimation angles.

\section{Conclusions}

We have calculated the radio light curves, and the evolution of the
apparent size and flux centroid (FC) of GRB jets viewed sideways.  As
the jets do not point at us, they will have no $\gamma$-ray precursor,
but will instead be preceded by a Type Ib/c SN.  A $\sim 1\;$yr old
GRB remnant at $D\sim 100\;$Mpc, is predicted to have a characteristic
radio flux of $\sim 100\;$mJy, an image size of $\sim 1\;$mas, and FC
motion of $\sim 20\;{\rm \mu{\rm as}~week^{-1}}$.  Such a source can
be resolved by VLBA at $D\lesssim 1\;$Gpc, while the motion of its FC
might be monitored up to $z\sim 1$ (although at $z\sim 1$ it would be
very dim, $\sim 0.1\;$mJy).  The apparent size of the jet or
super-luminal motion of its FC within the first few months after the
SN, may provide evidence for relativistic motion.

A relativistic jet of length $ct$ may also serve as a yardstick for
constraining cosmological parameters.  However, the required precision
for this purpose may not be attainable if the jet orientation is not
well known or the surrounding medium is inhomogeneous.

For an off-axis jet, there should be a time dependent linear
polarization, which peaks near the time of the peak in the light curve
and slowly decreases with time as the jet becomes more spherical and
symmetric around the line of sight (Granot et al. 2002).  If the jet
is resolved, then polarization maps could be generated, as is commonly
done for extragalactic radio jets.  This could reveal the magnetic
field geometry and orientation in the jet, and whether it has a large
scale ordered component (Granot \& K\"onigl 2003).

The existence of an early phase during which the emission of the jet
peaked in the UV ($\gamma \gtrsim 10$) can in principle be inferred from
the ionization cones preceding the jet in the surrounding gas (Perna
\& Loeb 1998).  Since the recombination time of the gas is $\sim 10^5
n_0^{-1}~{\rm yr}$, these cones should exist for long times after the
SN explosion. However, the separation between the ionization fronts
and the edges of the jet grows large only after the jet becomes
non-relativistic.  At these late times, one may detect emission lines
from highly ionized, metal rich gas (Perna, Raymond, \& Loeb 2000)
that reflect the hardness of the emission spectrum of the jet at
earlier times when it was highly relativistic. Detection of
ionization cones can be used to infer the early opening angle and
spectral flux of the jet at different frequencies (corresponding to
the the ionization state of different ions).  The latter can be used
to estimate the initial Lorentz factor of the jet, $\Gamma_0$ in the
range $\sim 5$--$20$, or determine if
$\Gamma_0\lesssim 5$ (no ionization cones), or $\Gamma_0\gtrsim 20$
(ionizing extending up to soft X-rays).

There is strong evidence connecting GRBs with SNe Ic that have a large
kinetic energy, $\gtrsim 10^{52}\;$erg (termed 'hypernovae' by
Paczynski 1998), and have a distinct spectral signature, as was
observed for SN 1998bw and SN 2003dh.  The search for GRB radio jets
in SN with such a spectrum is particularly interesting, as it could
show whether all such SNe produce GRBs. For $\theta_{\rm
  obs}<\theta_0$ the existence of relativistic jets can be revealed by
their high brightness temperatures ($\gg 10^{11}\;$K, e.g. Kulkarni et
al. 1998; Li \& Chevalier 1999).  The fraction of SN Ib/c that produce
GRB jets can help determine the local GRB rate and the distribution of
$\theta_0$. 

It may prove interesting to search for a correlation between the value
of $\Gamma_0$ as estimated from the ionization cones and the spectrum
of the SN. A correlation with ejecta energy and abundance patterns
(Maeda \& Nomota 2003) may show a continuous change in the SN spectrum
as a function of $\Gamma_0$ of the bipolar jets, which might indicate
that these jets are intimately related to, or perhaps are, the main
cause of the SN explosion.

\acknowledgments We thank Josh Winn, Mark Reid and Re'em Sari for
useful discussions.  This work was supported in part by the Institute
for Advanced Study, funds for natural sciences (J.G.) and NSF grants
AST-0071019, AST-0204514 \& NASA grant NAG 5-13292 (A.L.).  A.L.
acknowledges support from the IAS at Princeton and the J.S.
Guggenheim Memorial Fellowship.

\end{document}